\documentclass[10pt,letterpaper]{article}
\usepackage[top=0.85in,left=1.5in,right=1.5in,footskip=0.75in]{geometry}

\usepackage{amsmath,amssymb}
\usepackage{bm}

\usepackage{mathtools}

\usepackage{changepage}

\usepackage{ulem}

\usepackage{enumerate}

\usepackage[utf8x]{inputenc}

\usepackage{textcomp,marvosym}

\usepackage{cite}

\usepackage{nameref,hyperref}

\usepackage[right]{lineno}

\usepackage{microtype}
\DisableLigatures[f]{encoding = *, family = * }

\usepackage[table]{xcolor}

\usepackage{array}
 
\newcolumntype{+}{!{\vrule width 2pt}}

\newlength\savedwidth

\newcommand\thickhline{\noalign{\global\savedwidth\arrayrulewidth\global\arrayrulewidth 2pt}%
\hline
\noalign{\global\arrayrulewidth\savedwidth}}

\usepackage[aboveskip=1pt,labelfont=bf,labelsep=period,justification=raggedright,singlelinecheck=off]{caption}

\makeatletter
\renewcommand{\@biblabel}[1]{\quad#1.}
\makeatother

\usepackage{lastpage,fancyhdr,graphicx}
\usepackage{epstopdf}
\fancyheadoffset[L]{2.25in}
\fancyfootoffset[L]{2.25in}

\begin{document}
\vspace*{0.2in}

\begin{flushleft}
{\Large
\textbf{Predicting United States policy outcomes with Random Forests } 
}
\newline
\\
Shawn K. McGuire\textsuperscript{1*},
Charles B. Delahunt\textsuperscript{2*} 
\\
\bigskip
\textbf{1} Independent Researcher, Seattle, WA,
\\
\textbf{2} Applied Mathematics, University of Washington, Seattle, WA
\\ 
\bigskip
 
*smcguire@uw.edu,  delahunt@uw.edu

\end{flushleft}

%
%

\section{Abstract}

Two decades of U.S. government legislative outcomes, as well as the policy preferences of rich people, the general population, and diverse interest groups, were captured in a detailed dataset curated and analyzed by Gilens, Page \textit{et al.} (2014).
They found that the preferences of the rich correlated strongly with policy outcomes, while the preferences of the general population did not, except via a linkage with rich people's preferences.  
Their analysis applied the tools of classical statistical inference, in particular logistic regression.
In this paper we analyze the Gilens dataset using the complementary tools of Random Forest classifiers (RFs), from Machine Learning. 

We present two primary findings, concerning respectively prediction and inference:  
\textit{(i)} Holdout test sets can be predicted with approximately 70\% balanced accuracy by models that consult only  the preferences of rich people  and a small number of powerful interest groups, as well as policy area labels.  
These results include retrodiction, where models trained on pre-1997 cases predicted ``future" (post-1997) cases.   
The 20\% gain in accuracy over baseline (chance), in this detailed but noisy dataset,  indicates the high importance of a  few wealthy  players in U.S. policy outcomes, and  aligns with a body of research indicating that the U.S. government has significant plutocratic tendencies. \\ 
\noindent \textit{(ii)}  The feature selection methods of RF models identify especially salient subsets of interest groups (economic players).
These can be used to further
investigate the dynamics of governmental policy making, and also offer an example of the potential value of  RF feature selection methods for inference on datasets such as this.

%
%

\section{Introduction}

In 2014, Gilens and Page presented their ground-breaking paper “Testing Theories of American Politics: Elites, Interest Groups, and Average Citizens” \cite{originalGilens}, based on their research into the influence of various actors (power groups) on the outcome of important policy outcomes in the United States.  
The dataset spanned two decades, from 1981 to 2002, and consisted of policy cases determined to be of high importance.  
They identified and described an extensive array of independent variables  in their paper and in their book “Affluence and Influence: Economic Inequality and Power in America” \cite{gilensBook}. 

The two most important independent variables (hereafter ``features'') outlined in their work were: 

\textit{(i)} The 90\textsuperscript{th} income percentile's opinion (P90); 

\textit{(ii)} The net Interest Group Alignment (``netIGA''), a single derived variable combining the effects of 43 powerful interest groups (IGs).

They showed that public opinion at the 50\textsuperscript{th} percentile had little to no effect on the odds of policy adoption, challenging received notions of democracy in the United States.  

Additional research over the years has further questioned notions of American democracy.  
Thomas Ferguson’s research into the effect of major investors and money on election outcomes has shown that money flows are an excellent predictor of U.S. congressional races outcomes \cite{ferguson} \cite{goldenRule}.  
In 2016, the United States was downgraded by the UK-based Economic Intelligence Unit from``Full Democracy” to ``Flawed Democracy” \cite{revengeOfDeplorables}.  
Bartel's work has shown that the policy preferences of ``affluent" Americans correlate closely  with senators' roll-call votes and government policy \cite{unequalDemocracy}.

Gilens and Page applied classical methods of statistics and regression to analyse their dataset. 
Increasingly, Machine Learning (ML) methods are making their way into social science research \cite{mlForSociology}, and offer a distinct, complementary approach and set of tools for dataset analysis, one focused on prediction rather than inference \cite{breimanTwoCultures}.  

In the political sphere, ML methods have been applied to Supreme Court rulings \cite{katz}, and to passage of bills in the U.S. Congress based on text in the bills \cite{nay,yano, gerrish}.
Prediction does not require interpretability, and some ML methods are largely ``black-boxes'' (\textit{e.g.} neural nets).
But some ML methods (\textit{e.g.} Random Forests) have interpretable aspects, and are thus potentially useful for inference \cite{domingosAlgorithm}.

Logistic regression (used by Gilens and Page) includes an intrinsic, strong assumption of linearity \cite{logisticRegression} (for details see Appendix).  
Random Forests (RFs) in contrast are flexible, non-linear classifiers \cite{breimanRandomForests} that can handle large numbers of sparsely-represented features such as the preferences of the 43 individual IGs  in the Gilens dataset \cite{rfSparse}.
In addition,  RFs have natural metrics to assess the relative importance of each feature.
Thus RFs allow us to probe the dataset in complementary ways to Gilens and Page's use of a calculated netIGA  and logistic regression.

In this work we apply RF methods to the Gilens dataset with two goals, prediction and inference: 
We use RFs to build predictive models of U.S. policy; and we extend Gilens and Page's inferential findings as to the influence of various actors on U.S. policy outcomes. 
We offer two main findings: \\ 
(i) Policy outcomes on holdout sets can be predicted with approximately 70\% balanced accuracy (\textit{vs} 50\% chance baseline) using only a few feature categories from the Gilens dataset: Rich voters' preferences, a subset (as few as 14 out of 43) of individual IGs' preferences, and policy area labels. \\
(ii) The RF feature importance metrics enable further understanding and analysis of the salience of  individual actors, and also provide an example of RFs' potential usefulness for inference on datasets of this kind.

%
%

\section{Methods}

\subsection{Dataset}

The dataset consists of 1,836 major U.S. federal government policy cases dating from 1981 to 2002. 
In terms of prediction, the dependent variable $y$ is  case outcome (``adopted'' or ``not adopted''). We reduced the large array of features in Gilens' dataset to three primary categories, described below:  \textit{(i)} voter preferences (in particular, of the wealthy) \textit{(ii)} IG alignments, and \textit{(iii)} policy descriptors (such as ``Foreign'', ``Social Welfare'', \textit{etc}).  Corresponding abbreviations are listed in Table \ref{featureAbbreviations}.

\begin{table}[!ht]
\begin{adjustwidth}{0.2in}{0.2in} 
\centering
\caption{ \textbf{Feature abbreviations}
}
\begin{tabular}{|l|l| }
\hline
P90 & Voter preference of 90\textsuperscript{th} income \%ile \\ 
netIGA & Net Interest Group Alignment \\ 
IGs &  Individual Interest Groups (preferences) \\ 
PAs &  Policy Areas \\ 
PDs &  Policy Domains (a coarser version of PAs) \\ \hline
\end{tabular}
\begin{flushleft} 
\end{flushleft}
\label{featureAbbreviations}
\end{adjustwidth}
\end{table}

%
%

\paragraph{Voter preferences (P90):}
Preferences of different wealth tranches were obtained  from national surveys of the general public, where participants were asked whether they favored or opposed a  proposed policy change. 
Preferences of the different wealth tranches were then imputed at various income percentiles, viz 90\textsuperscript{th}, 50\textsuperscript{th}, and 10\textsuperscript{th} (hereafter P90, P50, P10). 
Gilens and Page noted that P90 was a (rough) indicator of the preferences of even higher income percentiles (e.g. P99).  
One of their key findings was that, among voter preferences, only P90 impacted case outcomes.  
Our models used P90 as the sole voter preference feature (use of P50 or P10 as features degraded model accuracy).  
%
%
 
\paragraph{Interest groups (IGs and netIGA):} 
The dataset includes alignments, on each case, for a total of 43 distinct IGs. 
These IGs  were mostly business groups such as the American Bankers Association, and also some social IGs such as the AARP (American Association of Retired Persons).
The alignment of each IG toward each case was assigned an ordinal value between 2 and -2 (Strongly Support, Somewhat Support, Neutral, Somewhat Oppose, or Strongly Oppose).
Routinely, a given IG had no opinion in a particular case, so non-neutral (non-zero) values were sparse. 

To aid their analysis, Gilens and Page combined the alignments of all the individual IGs to create a single feature, the ``netIGA'' as follows:
\begin{equation}
\label{netIGAEqn}
 \text{netIGA} = log(  F_2+   0.5 F_1  + 1) ~  
    - ~ log( O_2 +  0.5 O_1  +1), \text{ where}
\end{equation}  
$F_2$ = number of IGs strongly in favor, $F_1$ = number somewhat in favor, $O_2$ = number strongly opposed, and $O_1$ = number somewhat opposed.  
The $log(~)$ accounts for diminishing effects of multiple IGs weighing in on the same side of a given case.  

Because RFs readily handle large numbers of features, we set aside the netIGA and instead used each IG's alignment value as a separate feature.

%
%

\paragraph{Policy areas (PAs):}
The Gilens dataset  includes an in-depth policy descriptor category which we refer to as policy area (PA) labels, \textit{e.g.} Campaign Finance, Welfare Reform, \textit{etc}.    

A full list of the 19 PAs is given  in the Appendix.  
 PAs were expressed as features via one-hot encoding (\textit{i.e.} one feature per PA, taking values 0 or 1, according as the case was in that PA). 
Each case is assigned to exactly one PA.

\paragraph{Policy Domains (PDs):}
The Gilens dataset also includes six broader policy domain (PD) labels (Economic, Foreign, Social welfare, Religious, Guns, and Miscellaneous). In general, several PAs are contained in one PD, though some PAs are also PDs (\textit{e.g.} Foreign Policy).
PDs were one-hot encoded.
Table \ref{tablePosAndNegCountsByDomainAndDate} shows a breakdown of positive (\textit{i.e.} adopted) and negative (\textit{i.e.} not adopted) cases by PD over the full dataset, as well as over just the post-1997 test set (used for retrodiction).

PDs were used for feature selection:  
We grouped cases by PD, trained a different  RF model for each PD using P90 and individual IGs as features, then selected  the most salient IGs for each PD based on these models.
 
\begin{table}[!ht]
\begin{adjustwidth}{0.2in}{0.2in} 
\centering
\caption{ \textbf{Positive and negative case counts.}
Number of positive (``adopted'') and negative (``not adopted'') cases, as well as percentage of cases that were positive, for each PD, over the full dataset (numbers over post-1997 cases are in parentheses).
Economic policy and Foreign policy are fairly evenly-balanced, while other PDs skew negative.
In some PDs (\textit{e.g.} Social Welfare, Guns) the distribution of positive/negative cases differed pre- \textit{vs} post-1997. \\~ 
}
\begin{tabular}{|l|c|c|c| }
\hline
{\bf Domain}  & {\bf Pos} & {\bf Neg} & {\bf \%Pos}  \\ \hline
Economic & 160  (36) & 248 (38)  &  39\% (49\%)\\ \hline
Foreign & 244 (77)  &  196 (59) &  55\% (57\%)\\ \hline
Social Welfare&  101~(20)  & ~310 (152) & 25\% (12\%) \\ \hline
Religious & ~43 (18) & 123 (68)  &  26\% (21\%)\\ \hline
Guns & 18 (1)~ & ~81 (49)   & 18\% (2\%)~~\\ \hline
Misc &  ~77 (36) &  235 (95)   & 25\% (27\%)\\ \hline
Total & 643 (188) & 1193 (461)  & 35\% (29\%)\\ \hline
\end{tabular}
\begin{flushleft} 
\end{flushleft}
\label{tablePosAndNegCountsByDomainAndDate}
\end{adjustwidth}
\end{table}

\subsection{Train/test splits}

We trained models in two regimes:
\textit{(i)} Random train/test splits drawn from the full dataset ( \textit{N} = 25 draws, ratio 67\%, 33\%). 
This provided robust error bars on our prediction accuracy results since each train/test split was different.  
\textit{(ii)} Future prediction (more precisely, retrodiction): 
All pre-1997 cases in training and all post-1997 cases  (including 1997) in  testing (ratio 65\%, 35\%).  

Retrodiction tested stability over time, with the possibility that individual IGs might gain or lose relevance over the two decade duration of the dataset. 

\subsection{Reported metrics}
 
We report two figures of merit: maximum Balanced Accuracy, and Area Under the Curve (AUC) of the ROC curve.
Evaluated on the test set, these are standard measures of a model's classifying abilities. 
Maximum balanced accuracy is defined as
\begin{equation}
\label{balAccEqn}
balanced ~ accuracy = \frac{sensitivity + specificity}{2}, 
\end{equation} 

\noindent where sensitivity is the percentage of positive  cases correctly classified, and specificity  is the percentage of negative cases correctly classified.
The operating point (decision threshold) used to predict test cases is that which gives maximum balanced accuracy on the training set (this threshold typically gives lower test set accuracy than would be possible given oracle  knowledge of test set behavior).

We report balanced rather than raw accuracy for two reasons. 
First, it has a clear, consistent baseline accuracy (50\% = chance).
Second, in this dataset, negative cases outnumber positive ones, by substantial margins in some policy domains (cf. Table \ref{tablePosAndNegCountsByDomainAndDate}).
Given a class imbalance, raw accuracy blurs the differences between models because all models can leverage the imbalance by effectively betting on system inertia (\textit{i.e.} the fact that most legislation is not adopted). 
 
%
%

\subsection{Random Forest models}

RFs are flexible, robust, non-linear classifiers \cite{breimanRandomForests} based on ensembles of decision trees.
In a RF, many decision trees are generated, each of which trains on a random subset of the training data.
At each node of a given tree, a randomly-chosen feature splits the data.
The final prediction of a test case is an average of the trees' predictions of that case.
RFs are adept at handling sparse datasets with large number of features \cite{rfSparse}, though removing uninformative features can improve model accuracy.

Our code was written in Python \cite{pythonWebsite} and used the sklearn library \cite{sklearnWebsite}.
Full codebase can be found at \cite{githubCodebase}.

The accuracies of two other flexible classifiers, XGBoost \cite{xgBoost} (a flavor of RF) and Neural Nets,  were similar to  standard RFs (results not reported).  

\subsection{Feature selection}
To rank the importance of the various IGs as features, cases were divided into 6 groups according to Policy Domain. 
For each domain, RFs were trained on random train/test splits (\textit{N} = 21), using  P90 and the 43 IGs as features.
The averaged feature importance scores were then ranked.
For more details see the Appendix.

%
%

 
%
%


\section{Results}

Results are divided into two sections: Inference (feature selection by RFs); and Prediction (including retrodiction).

\subsection{Inference} 
\label{subsubsectionSelectionOfImportantIGs}
The goal of Inference was to identify the most salient IGs by ranking their power as predictors.
The  43 IGs were ranked as predictors for each Policy Domain as described above.
A typical feature ranking (for the Foreign Policy  domain) is shown in Table \ref{tableInterestGroupOutcomeCorrelationsForeign}.
Similar tables of IG rankings for other PDs are in the Appendix.

The  ranking  extracted those IGs with the most impact on policy outcomes, which has direct relevance to the study of policy decision-making in the U.S.
We note three points about the rankings:

(1) Saliency (defined as maximizing model accuracy) is not the same as effectiveness of influence, because an IG's success is also a function of the difficulty of the particular cases it lobbied for or against (by analogy, a baseball player's batting average is partly a function of the pitchers they face).

(2) Bias can be introduced into the rankings based on various predictor attributes \cite{biasRF}.
For example, less-sparse features (IGs with more at-bats) tend to have higher rankings.
In this dataset, restricting by Policy Domain mitigates this difference in sparsity since IGs tend  to be especially active in a particular PD.

(3) An IG can have predictive saliency due to negative correlations.

Correlations between IG preferences and case outcomes are given in Column 3 of Table \ref{tableInterestGroupOutcomeCorrelationsForeign}.
To calculate this correlation,  only cases where the IG was not neutral (\textit{i.e.} was ``at bat'') were considered: 

\begin{equation}
\label{igOutcomeCorrEqn}
corr(IG) =  \frac{0.5}{\#at~bats}\left( \sum\limits_{i | y_i = 1} x_i  ~ - \sum\limits_{i | y_i = 0} x_i \right), \text{  where }  
\end{equation}
$x_i$ =  IG preference $\in$\{ -2, -1, +1, +2\} and $y_i$ =  outcome $\in$\{ 0 or 1\},  for the \textit{i\textsuperscript{th}} case.
The 0.5 term normalizes the IG preference values. 
For these calculations, P90 values were rescaled from [0,1] to [-2, 2] and then values in [-0.4, 0.4] (i.e. noncommittal) were set to 0, to allow comparison with IGs.
 
\begin{table}[!t]
\begin{adjustwidth}{0.2in}{0.2in} 
\centering
\caption{ \textbf{Feature importance scores in Foreign policy.}
Feature importance scores, IG stance \textit{vs} case outcome correlations, and IG at-bats, given as mean $\pm$ std dev  over 25 random train/test splits of the full dataset, restricted to Foreign policy cases only.   
``at-bats'' gives the number of test cases for which the IG had a non-neutral stance (out of 147 $\pm$ 7 foreign policy test cases). 
}
\begin{tabular}{|l+c|c|c|c|c|c|}
\hline 
   & {\bf RF} &     {\bf IG-outcome }   &  {\bf  at-bats }  \\
  {\bf ~~~~~~~~IG} & {\bf  score} &   {\bf correlation }&  {\bf  (out of 147) }\\    
  \hline  
P90                            &   41 $\pm$  3   &  20 $\pm$   3      &  93 $\pm$ 6    \\ \hline
Defense industry  &   12 $\pm$   3     &   41 $\pm$  13    &  39 $\pm$ 5 \\ \hline
AIPAC                       & 7 $\pm$  2       &   -18 $\pm$  11   &  17 $\pm$ 4  \\ \hline
Auto companies  &    5  $\pm$  1     &   48 $\pm$  15    &  13 $\pm$   2\\ \hline
UAW union           &    5  $\pm$   1     &   -59 $\pm$  14   &  41 $\pm$   3\\ \hline
Oil companies    &   4 $\pm$   1         &   37  $\pm$  33  &   5 $\pm$  2 \\ \hline
Airlines                  &  4 $\pm$   1      &   54 $\pm$  23   &   8 $\pm$  2 \\   \hline  

\end{tabular} 
\begin{flushleft} Columns 2 and 3  are scaled by 100$\times$.  \\ 

\end{flushleft}
\label{tableInterestGroupOutcomeCorrelationsForeign}
\end{adjustwidth}
\end{table}

For example, in the Foreign policy domain (\textit{cf} Table \ref{tableInterestGroupOutcomeCorrelationsForeign}, Defense Contractors' stance was positively correlated with outcomes (mean $\pm$ std dev: 41 $\pm$ 13), AIPAC's stance had mixed correlation (18 $\pm$ 11), and Labor Unions' stances were negatively correlated (-59 $\pm$ 14).
Note that negative correlation does not mean that the IG ``lost'': Its actions may have averted worse outcomes, or modified the case's legislative details in desired ways.\\~
 
The most salient IGs for each Policy Domain, as determined by RF feature selection, were as follows:\\
1. Foreign policy: Defense contractors; then AIPAC, auto companies and auto workers union.\\
2. Economic policy: Construction and realtors; then tobacco companies and a union.\\
3. Social welfare policy: the AARP and investment companies; then governors, pharmaceutical companies, universities, teachers, and health insurance companies.\\
4. Religious policy: Christian and anti-abortion groups; then doctors, teachers, beer companies, tobacco companies, health insurance companies, and broadcasters.\\
5. Gun policy: the NRA (no other IG was active).\\
6. Miscellaneous: Chamber of Commerce; then two unions (AFL-CIO and government workers), automobile companies; then oil companies, Christian Coalition, and trial lawyers.\\
The preferences of the wealthy (P90) outranked every IG in every policy domain except Social Welfare, where the AARP dominated.

Tables of IG data  for each Policy Domain (feature importance, correlations with outcomes, and number of ``at-bats'') are given in the Appendix. 
 
\subsubsection{Advantages of RF's nonlinear flexibility in feature selection}
\label{nraVsP90}

RFs do not have the built-in linear structure of logistic regression. 
In the inference context, this flexibility means that RFs can give insights into feature salience which are unavailable to logistic regression.   
 
An example from the foreign policy domain data is described here. 
The highest ranked features for the RF model were P90 and Defense Contractors.
The logistic regression model for the foreign policy domain ranked P90 and Defense Contractors as the 2nd and 6th most salient features, respectively.  
We examined policy cases involving P90 and Defense Contractors to gain some comparative understanding.

\begin{figure}[!th] 
\begin{adjustwidth}{0.2in}{0.2in} 
\centering
{\includegraphics[width= 1\linewidth]{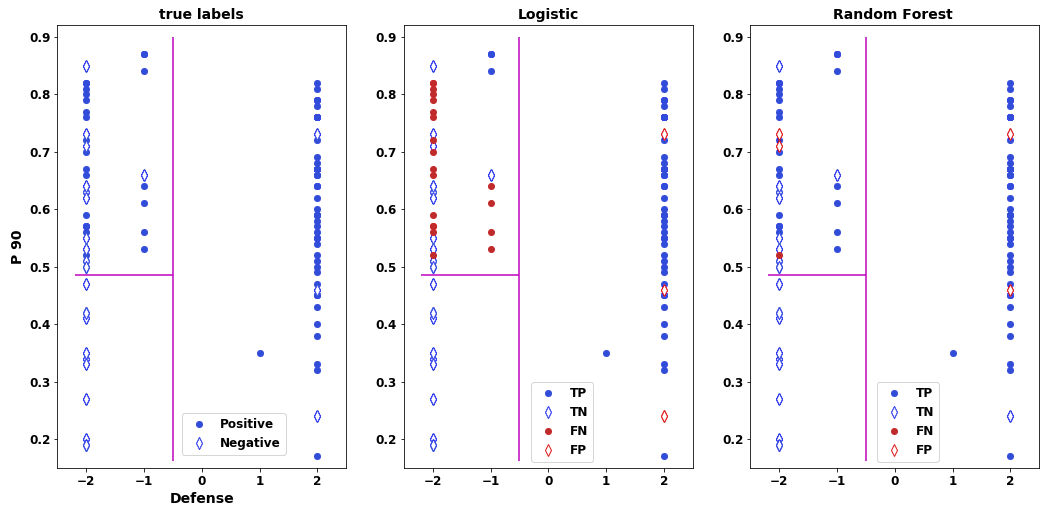}} 
\caption{{\bf Benefit of nonlinear splits.}
Foreign policy outcomes on cases where Defense Contractors had a non-neutral stance. 
\textit{y}-axis = P90, \textit{x}-axis = Defense Contractors' preference. 
Solid dots are positive outcomes, open diamonds are negatives.  TP = true positive, TN = true negative, FN = false negative, FP = false positive.
{\bf{left:}}  True outcomes.
There are three regions, marked by magenta lines:  When Defense Contractors favored a case it almost always passed, regardless of P90 (right half of plot). 
When Defense Contractors opposed a case then either:  it failed if P90 opposed it (lower left) or  it had reasonable odds of passing if P90 favored it (upper left). 
{\bf{middle:}} Logistic predictions (blue = correct, red = wrong), with many errors in the upper left region.
{\bf{right:}} RF predictions (blue = correct, red = wrong). 
RFs readily parse all three regions. }
 \label{figureRfSplitOfForeignCases}
\end{adjustwidth}
\end{figure}

When Defense Contractors strongly favor a policy change, they almost always get their way, even when the wealthy are opposed.  
However, when the Defense Contractors oppose a policy change, P90 has a strong positive correlation with outcomes. 
Logistic regression fails to parse this effect, giving many False Negatives (see Fig \ref{figureRfSplitOfForeignCases}). 
RFs handle this readily, yielding 95\% balanced accuracy for the RF \textit{vs} 79\% for logistic regression.  
That is, the high importance of P90 in Foreign policy cases depends on a relationship more readily encoded by the RF model. 

We note that RFs selected significantly more salient features than did logistic regression. 
For details see the Appendix, section \ref{sectionFeatureSelectionDetails}.

%
%
\subsection{Prediction} 

The goal of Prediction was to examine to what degree U.S. legislative outcomes might be predicted from the preferences of rich voters and Interest Groups, and which actors were most informative.

Our key finding is that legislative outcomes can be accurately predicted using P90, policy descriptors (PDs or PAs), and individual IGs.
Balanced accuracy on test sets of the trained RF reached $\approx$70\%, a gain of 20\% over baseline chance. 
 
P90 was a vital feature, in the sense that excluding it always degraded accuracy.
Inclusion of P50 (median wealth voter preferences)  as a feature slightly degraded accuracy, consistent with the finding of P50's irrelevance in \cite{originalGilens}.
Use of either PD or PA labels as a feature increased accuracy. 
Training and then combining separate models for each Policy Domain did not increase overall accuracy (results not reported).

We report results for RF models using four feature sets  (listed below), in two train/test scenarios: 
\textit{(i)} Multiple random train/test splits of the full dataset (``Random Draw'' in Table \ref{tableAccuracyOneModel}); 
and 
\textit{(ii)} Retrodiction (i.e., train on pre-1997 cases and test on post-1997 cases,  a single train/test split). 

Feature sets used were:

 {\bf Set A:} P90 and netIGA (Baseline, from \cite{originalGilens}).

 {\bf Set B:}  P90, PDs, and all 43 individual IGs.

{\bf Set C:}  P90, PDs, and 14 IGs chosen by RF feature selection (Gini impurity).

{\bf Set D:}  P90, PAs, and all 43 IGs.

Balanced accuracies and AUCs for RF models using the four Feature Sets and in the two regimes are given in Table \ref{tableAccuracyOneModel}.
We offer the following observations: 

(1) All feature sets used P90, which was always the most important. 
The feature sets differed in how they used IGs, and how they used Policy descriptors.

\begin{figure}[!ht] 
\begin{adjustwidth}{0.2in}{0.2in} 
\centering
{\includegraphics[width= 1 \linewidth]{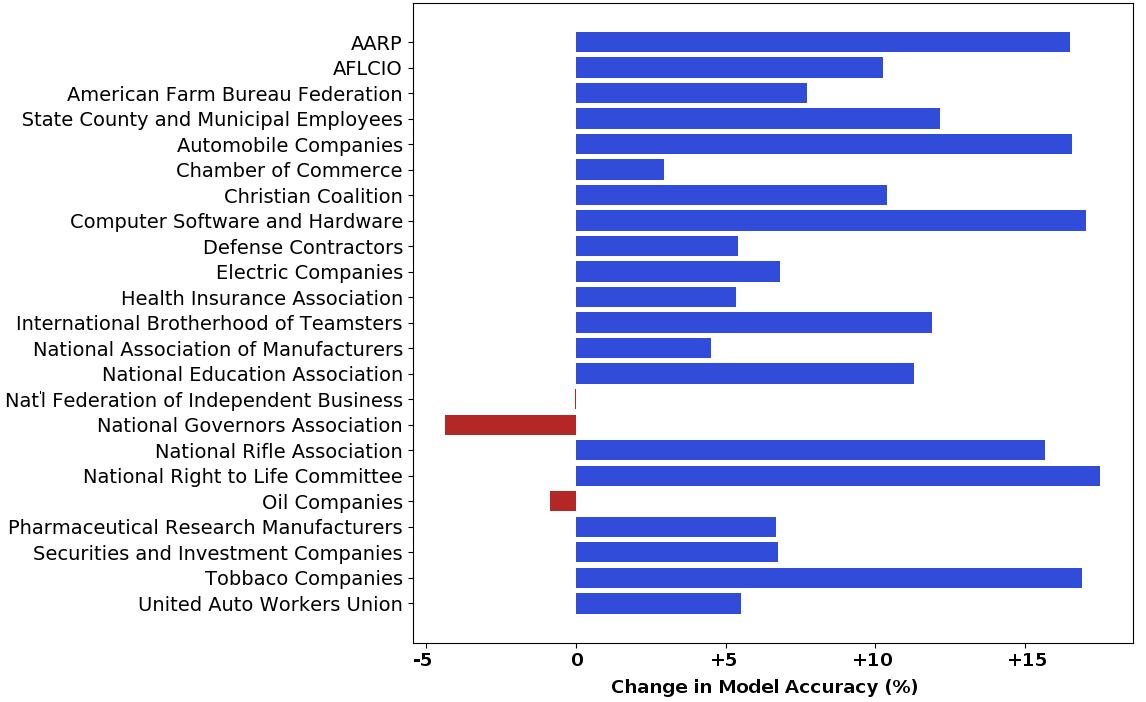}} 
\caption{{\bf Gain in accuracy per IG, Set B \textit{vs} Set A. }
Set B used P90, all 43 IGs and Policy Domains.  
Set A used  P90 and netIGA only.  
For each IG, we measured test set accuracy on those cases where the IG was ``strongly in favor'' or ``strongly opposed'' and plotted \textit{accuracy(Set B) - accuracy(Set A).} 
Set B consistently improved accuracy on these IG-based groups of cases.
Random draw regime, \textit{N} = 25. 
Each IG shown had at least 20 cases in the test set.}
 \label{figureRelAcc}
\end{adjustwidth}
\end{figure}

(2) Sets B - D substantially outperformed Set A (\textit{e.g.} 11\% mean increase in AUC).\\
Use of individual IGs \textit{vs} the simplified netIGA was the main factor behind this improvement in prediction accuracy.  
Figure  \ref{figureRelAcc} shows the mean improvement  in accuracy of Set B over Set A,  
broken out by individual IGs (``Random Draw'' regime).  
 
(3) Set C had equivalent performance to Set B, indicating that a small subset of 14 IGs (chosen by RF feature selection) carried as much salience as the full set of 43 IGs.
We note this does not imply that these 14 IGs were the only ones that mattered:
Certainly they were important actors, but they likely also encoded correlated salience of other IGs.  

(4) Set D posted better performance than Model B, indicating that Policy Area labels had more salience than the coarser Policy Domain labels. 

(5) Set D 
gave the best results (70\% balanced accuracy, 78\% AUC).    
The full list of features for Model D, with their importance rankings, is given in Table \ref{modelDFeatureImportance} in the Appendix.
 
 \begin{table}[!ht]
\begin{adjustwidth}{0.2in}{0.2in} 
\centering
\caption{ \textbf{RF accuracy with various feature sets.}
Balanced Accuracy and AUC for RFs given various feature sets.  
``PD'' means Policy Domains.
``PA" means Policy Areas.
Results are given as percentages, mean $\pm$ 1 std dev (\textit{N} = 25).}
\begin{tabular}{|l+c|c+c|c|} 
\hline 
 \multicolumn{1}{|c+}{\bf Model} & \multicolumn{2}{c+}{\bf Random Draw} & \multicolumn{2}{c|}{\bf Retrodiction}    \\ 
  & Bal Acc \% & AUC  \%& Bal Acc \% & AUC \%\\ \hline 

{\bf A:} P90, netIGA & 61.5 $\pm$ 1.7 & 66.2 $\pm$ 2.1 & 64.6 $\pm$ 0.6 & 70.2 $\pm$ 0.3\\ \hline
{\bf B:}  P90, 43 IGs, PDs & 67.3 $\pm$ 1.6 & 74.9 $\pm$ 1.7 &  69.0 $\pm$ 0.7 & 75.7 $\pm$ 0.2\\ \hline
{\bf C:}  P90, 14 IGs, PDs & 67.3 $\pm$ 1.6 & 75.1 $\pm$ 1.6 &  68.8 $\pm$ 0.5 & 75.7 $\pm$ 0.2\\ \hline
{\bf D:} P90, 43 IGs, PAs & {\bf 70.1  $\pm$ 1.5} & {\bf 77.7 $\pm$ 1.5} & {\bf 71.3 $\pm$ 0.7} &{\bf  76.5 $\pm$ 0.4}\\ \hline

\end{tabular}
\begin{flushleft} 
\end{flushleft}
\label{tableAccuracyOneModel}
\end{adjustwidth}
\end{table}


%
%


\section{Discussion}

In this work we applied the tools of Machine Learning to the Gilens dataset, as an aid to inference as well as for prediction.  
RF algorithms readily handle sparse datasets  that can be difficult for traditional regression techniques.  
RF methods enabled prediction of policy outcomes with surprising accuracy, especially considering the intrinsic noisiness of the dataset.
They also enabled us to examine the role of individual IGs and the effects of various features of the dataset. 
Additionally, we have shown  how RFs can provide insights into  feature saliency distinct from and arguably superior to traditional regresssion techniques such as logistic regression. 
 
We found that case outcomes on test sets were predictable with $\approx$70\% balanced accuracy 
using only the preferences of the wealthy (P90), individual Interest Groups, and Policy Area labels as features.
Inclusion of the wishes of median income voters (P50) always reduced predictive accuracy.  
These results also held for a retrodiction split,  training  on pre-1997 cases and testing on post-1997 cases.
We believe that the high predictability of policy outcomes using models based on only a few wealthy actors (P90 and certain IGs) reinforces Gilens and Page's findings about the plutocratic tendencies of U.S. government. 

We suspect that the 90\textsuperscript{th} percentile serves as a proxy for the opinions of the very wealthy, and we expect that  predictive accuracy would further  improve given better measures of the opinions of the ``very wealthy” (\textit{e.g.} the 99\textsuperscript{th} income percentile).  
 
There has been effort in recent years to capture the policy opinions of the very wealthy as distinct from the general population.
A unique  study led by Benjamin Page   effectively captured, via surveys and interviews, the policy preferences of those in, or near, the top 1\% of income in the United States \cite{policyPreferencesOfWealthy}.  
The results indicated, among other things, that the very wealthy held policy preferences that were much more conservative in such important domains as social welfare programs, taxation, and regulation of the economic system.  
Their research also showed that the very wealthy were much more politically active, with significantly higher rates of financial contributions to, and contact with, public officials.  
This highlights the importance of  quantitatively capturing the opinions of the very wealthy in the future, in order to better understand their impact on policy adoption by the U.S. government. 
 
%
%


\section{Appendix}

\subsection{Logistic regression}
Given a feature-label pair \{$\bm{x}, y$\}, logistic regression gives an  estimate $\hat{y}$ of the class $y$, using features $\bm{x}$, by passing a linear combination of feature values $\bm{\beta x}$ through a (non-linear) sigmoid function $ \sigma(s) = \frac{1}{1 - e^{-s}}$. The coefficients $\bm{\beta}$ are the fitted parameters. 
\begin{equation}
\hat{y} = \frac{1}{1 - e^{-\bm{\beta x}}}
\end{equation} 

\subsection{Feature selection details}
\label{sectionFeatureSelectionDetails} 

In order to identify the most important IGs, we used RF's standard Gini impurity method \cite{breimanClassification}, indicated for this dataset  because the features were sparse, non-categorical, and similarly scaled (in range [-2, 2]) \cite{rfSparse} (because IGs were ranked for each Policy Domain separately, the PDs were not features.)

For feature ranking purposes, P90 scores were rescaled from [0,1] to [-2, 2]  to match the IG preference range.
Unlike Gini impurity, RF's Permutation method had unstable results (IG rankings changed if P90 was excluded).
In general, the RFs selected different subsets of important IGs than those selected via the $\beta$ coefficients of logistic regression.

Features selected by RF (Gini impurity) significantly out-performed features selected by logistic regression, in the sense that  using RF-Gini features gave a model (either RF or logistic regression) much higher accuracy than  using features selected by logistic regression.   
Features chosen by FR's Permutation method landed between the two.
Table \ref{tableRfVsLogisticChosenFeatures} shows the gain in accuracy due to RF (Gini)-chosen features vs logistic-chosen features.

We note that for a fixed set of features, RF and logistic regression models gave similar accuracies (in general, RF accuracy was slightly but not significantly better, consistent with \cite{rfVsLogistic1}).
The key advantage of RFs lay not in prediction \textit{per se}, but rather in selecting much more salient features (\textit{i.e.} inference).

\begin{table}[!ht]
\begin{adjustwidth}{0.2in}{0.2in} 
\centering
\caption{ \textbf{Difference in salience of IGs chosen by RF \textit{vs} logistic regression.}
Balanced accuracy (balAcc) and AUC of RF and logistic models on test sets, using either RF-chosen IGs or logistic-chosen IGs, given as mean $\pm$ std dev. 
The gains in accuracy (RF-chosen minus Logistic-chosen) are shown in bold. 
(\textit{N}= 21 train/test splits). 
}
\begin{tabular}{|c|c| c|c|c|c| }
\hline \multicolumn{2}{|c|}{\bf Setup} & \multicolumn{2}{c|}{\bf Full data} & \multicolumn{2}{c|}{\bf Retrodiction}    \\
  { Model} & {  IG type} &    { balAcc} & { AUC} &   { balAcc} & { AUC }\\  \thickhline
 &  RF-chosen (Gini)              &  63.5  $\pm$ 1.6         &   68.2  $\pm$ 2.2         &     66.4   $\pm$ 2.4      & 69.7   $\pm$ 1.0      \\ RF   &  Logistic-chosen     &  56.7    $\pm$ 1.7        &  61.1  $\pm$ 1.9         &      53.4    $\pm$  2.7   &     53.9 $\pm$  2.7          \\ 
 & Gain  &    \bf 6.8  $\pm$ 1.9   &    \bf 7.0  $\pm$ 1.6  &  \bf 13.0    $\pm$ 4.4 &   \bf  15.8    $\pm$   3.3      \\ \hline
& RF-chosen (Gini)      &  59.5   $\pm$ 1.8         &  61.4   $\pm$ 2.1        &    59.3                             &  62.6         \\ 
Logistic   & Logistic-chosen   &  55.0     $\pm$ 1.6 &  56.6    $\pm$ 1.6      &   52.6                             &   51.6       \\ 
& Gain &  \bf 4.5  $\pm$ 1.9 & \bf 4.7  $\pm$ 1.7 &  \bf  6.7                         &   \bf  10.9      \\ \hline
\end{tabular}
\begin{flushleft}
``Gain'' is the mean of differences (not difference of means).\\
Logistic regression is deterministic on non-separable data, so its prediction of the post-1997 data  has 0 std dev.\\
\end{flushleft}
\label{tableRfVsLogisticChosenFeatures}
\end{adjustwidth}
\end{table}

\subsection{IG rankings by Policy Domain} 

Importance rankings,  correlations with case outcomes, and number of ``at-bats'' for IGs, for various Policy Domains, are given  in Tables  \ref{tableInterestGroupOutcomeCorrelationsEconomic} - \ref{tableInterestGroupOutcomeCorrelationsGuns}.
For more details, and for a table of Foreign policy results, see section \ref{subsubsectionSelectionOfImportantIGs}.
     
\begin{table}[!h]
\begin{adjustwidth}{0.2in}{0.2in} 
\centering
\caption{ \textbf{Economic policy.}
Correlations  for the most salient IGs, given as mean $\pm$ std dev. 
Each test set had 130 $\pm$ 6  Economics  cases.
}
\begin{tabular}{|l+c|c|c|c|c|c|}
\hline 
   & {\bf RF} &   {\bf IG-outcome }   &  {\bf  at-bats }  \\
  {\bf ~~~~~~~~IG} & {\bf  score} &  {\bf correlation }&  {\bf  (out of 130) }\\    
  \hline  
P90                                 &   21 $\pm$  2       &   12 $\pm$   4      &  94 $\pm$ 5    \\ \hline
Nat'l Assoc Homebuilders            &      6  $\pm$  2       &   23 $\pm$  7       &  41 $\pm$ 5 \\ \hline
Nat'l Assoc Realtors                        &   6 $\pm$  2       &   46 $\pm$   11    &  22 $\pm$ 4  \\ \hline
Tobacco Companies  &    5  $\pm$  1    &   18 $\pm$  11     &  31 $\pm$   6\\ \hline
Teamsters union         &    5  $\pm$   1       &   40 $\pm$  11   &  19 $\pm$   3\\ \hline

\end{tabular}
\begin{flushleft} Columns 2 and 3 are scaled by 100$\times$. 
\end{flushleft}
\label{tableInterestGroupOutcomeCorrelationsEconomic}
\end{adjustwidth}
\end{table}

    
\begin{table}[!h]
\begin{adjustwidth}{0.2in}{0.2in} 
\centering
\caption{ \textbf{Social Welfare policy.}
Correlations  for the most salient IGs, given as mean $\pm$ std dev. 
The AARP had strong positive correlation with outcomes, while 90\textsuperscript{th} \%ile was mixed and investment interests had negative correlations.
Each test set had 137 $\pm$ 7  Social Welfare cases.}
\begin{tabular}{|l+c|c|c|c|c|c|}
\hline 
   & {\bf RF} &     {\bf IG-outcome }   &  {\bf  at-bats }  \\
  {\bf ~~~~~~~~IG} & {\bf  score} &   {\bf correlation }&  {\bf  (out of 137) }\\    
  \hline  
AARP                            &   23 $\pm$  2    &  60 $\pm$   8     &  71 $\pm$ 7    \\ \hline
P90                               &   18  $\pm$  2   &     -3 $\pm$  5     &  88 $\pm$ 9 \\ \hline
Invest \& Securities Assoc    & 13 $\pm$  2  &   -94 $\pm$   8    &  15 $\pm$ 3  \\ \hline
Nat'l Governors Assoc &    6  $\pm$  1 &     4 $\pm$  20    &  22 $\pm$   4\\ \hline
Pharmaceuticals      &    5  $\pm$   1     &   48 $\pm$  22    &  13 $\pm$   4\\ \hline
Universities               &  5 $\pm$   2    &  55  $\pm$ 50   &  3  $\pm$  1  \\ \hline
Nat'l Education Assoc          &  4 $\pm$    1      & 27   $\pm$ 25   &  16  $\pm$ 3   \\ \hline
Health insurance Assoc    &  4 $\pm$   1        & 36   $\pm$ 22   &  21  $\pm$  3  \\ \hline

\end{tabular}
\begin{flushleft} Columns 2 and 3 are scaled by 100$\times$. 
\end{flushleft}
\label{tableInterestGroupOutcomeCorrelationsSocialWelfare}
\end{adjustwidth}
\end{table} 
 
    
\begin{table}[!h]
\begin{adjustwidth}{0.2in}{0.2in} 
\centering
\caption{ \textbf{Religious policy.}
Correlations  for the most salient IGs, given as mean $\pm$ std dev. 
Only 12 IGs had non-zero alignments on Religious cases, and only 2 IGs were often non-zero.
Most features, including P90, had mixed correlations with outcomes.
Anti-abortion groups had negative correlations.
Each test set had 56 $\pm$ 5 Religious cases.
}
\begin{tabular}{|l+c|c|c|c|c|c|}
\hline 
   & {\bf RF} &   {\bf IG-outcome }   &  {\bf  at-bats }  \\
  {\bf ~~~~~~~~IG} & {\bf  score} &  {\bf correlation }&  {\bf  (out of 56) }\\    
  \hline  
P90                                      &   36 $\pm$  3       &  6  $\pm$   5        &  30 $\pm$ 4      \\ \hline
Nat'l Right to Life           &   23  $\pm$  2    &   -55 $\pm$  19   &  16 $\pm$ 4   \\ \hline
Christian Coalition          & 10 $\pm$  2     &   -14 $\pm$   14  &  43 $\pm$ 4    \\ \hline
Am. Medical Assoc          &    7  $\pm$  2     &   -21 $\pm$  38   &  4 $\pm$   1    \\ \hline
Nat'l Education Assoc    &   6  $\pm$   2    &   65 $\pm$  26   &  5 $\pm$   1    \\ \hline
Beer Companies             &  4 $\pm$   1        &  -34  $\pm$ 72   &  2  $\pm$  1    \\ \hline
Tobacco Companies      &  4 $\pm$    1     & -11   $\pm$ 55   &  4  $\pm$ 1    \\ \hline
\end{tabular}
\begin{flushleft} Columns 2 and 3 are scaled by 100$\times$.  
\end{flushleft}
\label{tableInterestGroupOutcomeCorrelationsReligion}
\end{adjustwidth}
\end{table} 

\begin{table}[!h]
\begin{adjustwidth}{0.2in}{0.2in} 
\centering
\caption{  \textbf{Gun policy.}
Correlations   for the most salient IGs, given as mean $\pm$ std dev. 
The NRA was the only active IG, and it had strong positive correlation with outcomes.
P90 (and P50) had low overall  correlation with outcomes, but paradoxically had much higher importance scores (both RF and Logistic) than did the NRA.
P90's high RF importance score was likely due to its non-linear splitting ability even while it did not ``get what it wanted'' (similar to the situation described in section \ref{nraVsP90}).
Each test set had 33 $\pm$ 6 Gun cases.
}
\begin{tabular}{|l+c|c|c|c|c|c|}
\hline 
   & {\bf RF} &   {\bf IG-outcome }   &  {\bf  at-bats }  \\
  {\bf ~~~~~~~~IG} & {\bf  score} &   {\bf correlation }&  {\bf  (out of 33) }\\    
  \hline  
P90                            &   96 $\pm$  1    &  -11 $\pm$   11      &  28 $\pm$ 5    \\ \hline
Nat'l Rifle Assoc  &   4  $\pm$  1      &   51 $\pm$  16    &  32 $\pm$ 5 \\ \hline
 
\end{tabular}
\begin{flushleft} Columns 2 and 3 are scaled by 100$\times$.   
\end{flushleft}
\label{tableInterestGroupOutcomeCorrelationsGuns}
\end{adjustwidth}
\end{table} 

\clearpage

\subsection{Set D feature importances.}
Feature Set D gave the most accurate predictions. 
Features consisted of: 
P90 (continuous values $\in$ [0,1]); 43 individual interest groups (values $\in$ \{-2, -1, 0, 1, 2\}); and 19 Policy Areas (Budget, Campaign Finance, Civil Rights, Defense, Economics and Labor, Education, Environment, Foreign Policy, Government Reform, Guns, Health, Immigration, Miscellaneous, Race, Religion, Social Welfare, Taxation, Terrorism, Welfare Reform, all one-hot encoded).
Table \ref{modelDFeatureImportance} shows feature rankings found by RFs (Gini impurity) for Set D.
Note that because this feature set combines one-hot and ordinal value encodings, the caveats in \cite{rfSparse} and \cite{biasRF} may apply.

\begin{table}[!ht]
\begin{adjustwidth}{0.2in}{0.2in} 
\centering
\caption{ \textbf{Set D feature importances.}  
Mean scores (\textit{N} = 50 runs) of the top 25 most salient features.  
Policy Area features indicated by  ``(PA)'' before the feature.
RF importance scores are scaled by 100$\times$.
}
\begin{tabular}{|l|c| }
\hline
	\bf Feature & \bf Mean RF score \\ \hline
	P90 & 27.3 \\ \hline
	AARP & 8.9 \\ \hline
	(PA) Foreign Policy & 7.9 \\ \hline
	Defense Contractors & 4.9 \\ \hline
	Chamber of Commerce & 2.8 \\ \hline
	(PA) Social Welfare & 2.1 \\ \hline
	National Association of Realtors & 2 \\ \hline
	National Right to Life Committee & 2 \\ \hline
	Health Insurance Association & 1.8 \\ \hline
	Christian Coalition & 1.8 \\ \hline
	National Rifle Association & 1.7 \\ \hline
	(PA) Health & 1.7 \\ \hline
	AFLCIO & 1.5 \\ \hline
	(PA) Campaign Finance & 1.5 \\ \hline
	(PA) Defense & 1.4 \\ \hline
	American Farm Bureau Federation & 1.3 \\ \hline
	American Israel Public Affairs Committee & 1.3 \\ \hline
	(PA) Guns & 1.2 \\ \hline
	Securities and Investment Companies & 1.2 \\ \hline
	National Federation of Independent Business & 1.2 \\ \hline
	National Association of Manufacturers & 1.1 \\ \hline
	Automobile Companies & 1.1 \\ \hline
	United Auto Workers Union & 1 \\ \hline
	Tobbaco Companies & 1 \\ \hline
\end{tabular}
\begin{flushleft} 
\end{flushleft}
\label{modelDFeatureImportance}
\end{adjustwidth}
\end{table}


\nolinenumbers

%
%
%

\end{document}